\documentclass[aps,pre,twocolumn,showpacs,floatfix,superscriptaddress]{revtex4-1}
\usepackage{graphicx}
\usepackage{graphics}
\usepackage[usenames]{color}

\newcommand{\be}{\begin{equation}}
\newcommand{\ee}{\end{equation}}
\newcommand{\bea}{\begin{eqnarray}}
\newcommand{\eea}{\end{eqnarray}}

\begin{document}
\title{Shock propagation in locally driven granular systems}
\author{Jilmy P. Joy}
\email{jilmyp@imsc.res.in}
\affiliation{The Institute of Mathematical Sciences, CIT Campus, Taramani, 
Chennai-600113, India}
\affiliation{Homi Bhabha National Institute, Training School Complex, Anushakti Nagar, Mumbai 400094, India}
\author{Sudhir N. Pathak}
\email{snpathak@ntu.edu.sg}
\affiliation{Division of Physics and Applied Physics, School of Physical and Mathematical Sciences, Nanyang Technological University, 
Singapore}
\author{Dibyendu Das}
\email{dibyendu@phy.iitb.ac.in}
\affiliation{Department of Physics, Indian Institute of Technology, Bombay, Powai, Mumbai-400076, India}
\author{R. Rajesh}
\email{rrajesh@imsc.res.in}
\affiliation{The Institute of Mathematical Sciences, CIT Campus, Taramani, 
Chennai-600113, India}
\affiliation{Homi Bhabha National Institute, Training School Complex, Anushakti Nagar, Mumbai 400094, India}
  
\date{\today}

\begin{abstract}
We study shock propagation in a system of initially stationary hard spheres that is driven by a continuous
injection of particles at the origin. The disturbance created by the injection of energy spreads radially
outward through collisions between particles.  Using scaling arguments, we determine the
exponent characterizing the power law growth of this disturbance in all dimensions. The scaling functions
describing the various physical quantities are determined using large-scale event-driven simulations in two and three dimensions for both elastic and inelastic systems.   
The results are shown to describe well the data from two different experiments on granular systems 
that are similarly driven.
\end{abstract}

\maketitle

\section{\label{introduction} Introduction}

Granular materials are  ubiquitous in nature. Examples include geophysical flows \cite{cscampbell1990arfm}, large-scale structure  formation of the 
universe~\cite{sfshandarin1989rmp}, sand dunes~\cite{hnishimoriprl1993}, craters~\cite{amwalsh2003prl}, etc. The  dissipative nature of the  
interactions among the constituent particles can lead to diverse physical phenomena such as pattern formation, clustering instability, 
granular piles, jamming, segregation, stratification, shear flows,  surface waves, fingering instability, and fluidization (see the reviews in~\cite{hmjaeger1996rmp,isaranson2006rmp,lpkadanoff1999rmp}). 
A subclass of problems that have been of experimental and theoretical interest is the response of a granular system at rest to an external perturbation
that is applied either as an instantaneous impulse or continuously in time. This phenomenon has been studied in many different
contexts, examples of which include avalanches in sand piles as a response to the addition of sand grains~\cite{adaerr1999nature}, 
crater formation on granular beds due to the impact of an external object~\cite{ptmetzger2009aip,jfboudet2009prl}, growing craters due to impinging jets on 
granular piles~\cite{ygrasselli2001granmatt}, shock formation in flowing granular media due to external impact~\cite{jfboudet2009prl}, viscous 
fingering due to constant injection of
particles~\cite{xcheng2008natphy,bsandnes2007prl,sfpinto2007prl,ojohnsen2006pre,hhuang2012prl}, {and formation of bastwaves in astrophysical systems~\cite{jpostriker1988rmp}. The externally applied perturbation often results in a disturbance that  grows in time as a power law and the power-law exponents may often be obtained by studying simple tractable models of suitably excited  spherical particles  where energy dissipation is only through  
inelastic collisions~\cite{snpathak2012pre}. We discuss below
the  response to perturbation in the context of such models.

One of the most commonly studied examples is the {\it globally perturbed} freely cooling granular gas, where homogeneously distributed 
macroscopic particles with random initial velocities move ballistically and dissipate energy through inelastic collisions,
in the absence of any external driving. Here the perturbation is the energy that is initially given. In the early stage of evolution, 
when the system is spatially homogeneous, kinetic energy of the system 
$E(t)$  decays with time $t$ as $t^{-2}$ (Haff's law)~\cite{pkhaff1983jfm} in all dimensions. 
At later times, due to inelastic collisions, the system becomes spatially inhomogeneous~\cite{igoldhirsch1993prl,smcnamara1996pre} and energy 
decreases as $t^{-\theta_{d}}$, where $\theta_{d}$ is less  than 2 and depends on dimension $d$ \cite{ebennaim1999prl,mshinde2007prl,mshinde2009pre,mshinde2011pre,lfrachebourg1999prl,schen2000phylettA,
smiller2004pre,xnie2002prl,snpathak2014epl,snpathak2014prl}. Haff's law for the homogeneous regime has been confirmed in experiments~\cite{ccmaab2008prl,statsumi2009jfm}, 
while $\theta_{d}$ characterizing the inhomogeneous regime has still not been observed in any experiment.

A different limit is the {\it locally perturbed} freely cooling granular gas, where initially all particles are at rest and kinetic energy is imparted to a few 
localized particles. Due to collisions, the disturbance grows radially outward, with a shock front separating the moving particles from the 
stationary ones. The elastic version of this problem has great similarity to the problem of shock propagation following an intense explosion.  
The hydrodynamic description of the 
propagation in a conservative fluid is the famous Taylor-von Neumann-Sedov (TvNS) solution~\cite{gtaylor1950,lsedov_book,jvneumann1963cw}. This solution is relevant in the experimental studies of the production of a cylindrically symmetric blast wave produced by ultrafast laser pulses~\cite{mjedwards2001prl}. Numerical simulations of the elastic system are consistent with the 
TvNS exponents~\cite{tantal2008pre,zjabeen2010epl}. In the inelastic system, the disturbance is concentrated in dense bands 
that move radially outward, and the relevant
exponents may be obtained through scaling arguments based on the conservation of radial momentum~\cite{zjabeen2010epl,snpathak2012dae,snpathak2012pre}. 
The variation of physical quantities inside the dense band may be obtained through a hydrodynamic description~\cite{mbarbier2015prl,mbarbier2015jstatphys}. The exponents obtained thus may be used to describe~\cite{snpathak2012pre} 
experiments on shock propagation in flowing glass beads that are perturbed by the  impact of steel balls~\cite{jfboudet2009prl}.

In both cases discussed above, the perturbation was an impulse. One could also consider {\it continuous and locally perturbed} driven granular
systems, where particles at rest are driven by a continuous injection of energy in a small domain. This scenario has been investigated
in many recent experiments and  includes pattern  formation  in  granular material due to the injection of a gas~\cite{xcheng2008natphy,ojohnsen2006pre}, grains~\cite{sfpinto2007prl},  or fluid ~\cite{hhuang2012prl}. There is currently no model that determines the exponents for such situations. In this paper
we study a simple model of spheres at rest that is driven at the origin by a continuous injection of particles from outside. From a combination
of event-driven simulations and scaling arguments, we determine the exponents governing the growth of the disturbance. The results
are compared with the data from two experiments~\cite{xcheng2008natphy,ojohnsen2006pre} and excellent agreement is obtained.

The remainder of the paper is organized as follows. 
In Sec.~\ref{model} we define the model precisely and give details of the event-driven simulations that we performed. 
The exponents characterizing the growth of the different physical quantities in the problem are determined using scaling
arguments in Sec.~\ref{scalingargument}. The assumptions and predictions of the scaling argument are tested using 
large-scale simulations in Sec.~\ref{numericresult} for both the elastic and the inelastic system. 
In Sec.~\ref{comparisonwithexperiments} we show that the results in this paper are able to explain data
from two experiments on driven granular systems.  Section~\ref{conclusion} contains a brief summary and a discussion of results.

\section{\label{model}Model}

Consider a $d$-dimensional system of hard spheres whose mass and diameter are set to one. The particles move ballistically until they undergo momentum-conserving binary collisions with other particles. If $\vec{u}{_1}$ and $\vec{u}{_2}$ are the velocities of two particles $1$ and $2$ before collision, then the velocities after collision, $\vec{v}{_1}$ and $\vec{v}{_2}$, are given by
\bea
\vec{v}{_1} &=& \vec{u}{_1} - \frac{1+r}{2} [\hat{n} \cdot (\vec{u}{_1}-\vec{u}{_2})] \hat{n},\\
\vec{v}{_2} &=& \vec{u}{_2} - \frac{1+r}{2} [\hat{n} \cdot (\vec{u}{_2}-\vec{u}{_1})] \hat{n},
\eea
where $r$ is the coefficient of restitution and $\hat{n}$ is the unit vector along the line joining the centers of particles $1$ and $2$. In a collision, the tangential component of the relative velocity remains unchanged, while the magnitude of the longitudinal component is reduced by a factor $r$. 
The collisions are elastic when $r=1$, and inelastic and dissipative otherwise.

Initially, all particles are at rest and uniformly distributed in space. The system is  driven locally by a continuous input of energy restricted to a small region
by injecting particles at a constant rate $J$ at the origin.  The injected particles have a speed $v_0$ in a randomly chosen direction until they undergo their first collision, after which the injected particles are removed from the system. Driving in this manner injects energy into the system, but conserves the total
number of particles. We will refer to this model as the conserved model.

We also consider a nonconserved model. This model is identical to the conserved model described above, but the injected particles stay in the system,
thereby increasing the total number of particles at a constant rate $J$. While the conserved model is applicable to two-dimensional granular systems driven
by a gas (where the gas may escape in the third dimension), the non-conserved model is applicable to two-dimensional granular systems driven
by granular material. We will show in Sec.~\ref{scalingargument} that the scaling laws at large times are identical for both models. We will therefore present numerical results only for the conserved model.

We simulate systems with number density 
$0.25$ (packing fraction $0.196$) in two dimensions and $0.40$  (packing fraction $0.209$) in three dimensions,  using event-driven molecular dynamics~\cite{dcrapaportbook}. 
These number densities are much smaller than the random close-packed density. The total number
of particles is $8 \times 10^6$ and is large enough such that the disturbance induced by the injection of particles does not  reach the boundary up to the simulation times considered in this paper.
We set  $v_0=1$, the rate of injection of particles $J$ is set to $1$, and  the injected particles have the same mass and diameter as the other particles in the system. 
In the simulations, the collisions are inelastic with constant restitution coefficient $r$ when the relative velocities of the particles are
greater than a cut off velocity $\delta$ and considered to be elastic otherwise. This procedure prevents the occurrence of the inelastic
collapse of infinite collisions within a finite time, which is a hindrance in simulations,  and is also  in accordance with the fact that
the coefficient of restitution tends to $1$ with decreasing relative velocity between the colliding particle~\cite{cvraman1918pr}.
The value of $\delta$ is $10^{-4}$, unless specified otherwise. 
The results are independent of $\delta$.

The numerical results in this paper are shown only for the conserved model and are typically averaged over $48$ different 
realizations of the initial particle configurations. All lengths are 
measured in units of the particle diameter and time in units of initial 
mean collision time $t_0=v_0^{-1} n^{-1/d}$, where $n$ is the number density.

\section{\label{scalingargument}Scaling Argument}

In order to develop scaling arguments to describe the propagation of energy, it is important to first visualize how the inelastic system  evolves 
in comparison to the elastic system. When the energetic particles are injected from the center, in both cases particles get disturbed up to a distance and the zone of disturbance propagates radially outward. Figures~\ref{fig:snapshotselastic} and \ref{fig:snapshotsinelastic} show the time evolution of the elastic and inelastic systems with $r=1$ and $r=0.1$, respectively, in two dimensions for the conserved model.  
In the elastic system, the circular region of moving particles (marked in red) has nonzero density everywhere.  In contrast, in the case of the inelastic system, particles  cluster together and form a dense band adjacent to the front of the
disturbance, forming a vacant region around the center. This circular band  moves outward with time and grows by absorbing more particles. We
observe the  same features in the simulations of the nonconserved model.
\begin{figure}
\includegraphics[width=\columnwidth]{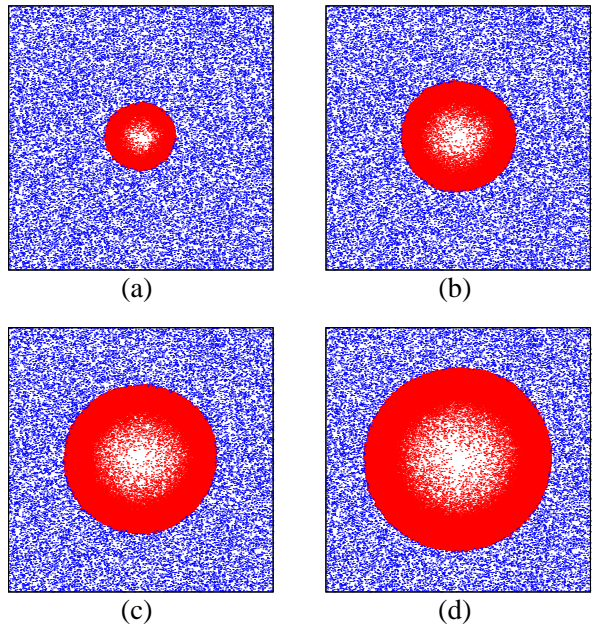}
\caption{(Color online) Moving (red) and stationary (blue) particles at times (a) $t = 500$, (b) $t = 1000$, (c) $t = 1500$ and  (d) $t = 2000$. Energetic particles are 
injected at the center. All collisions are elastic with $r=1$. The data are for the conserved model.}
\label{fig:snapshotselastic}
\end{figure}
\begin{figure}
\includegraphics[width=\columnwidth]{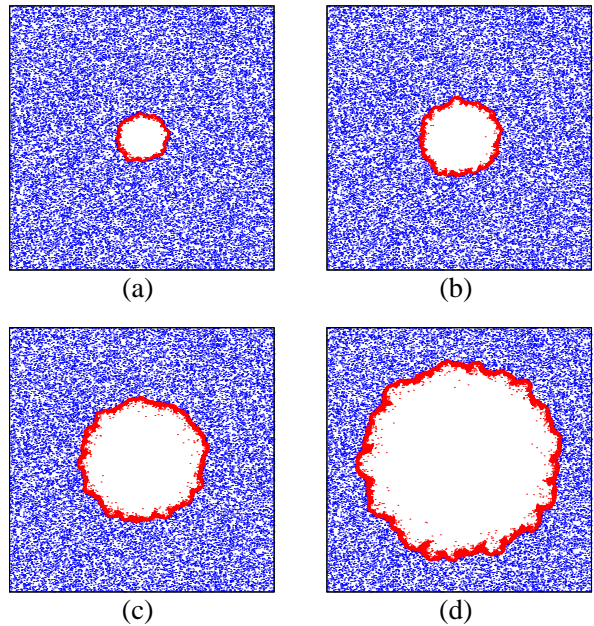}
\caption{(Color online)  Moving (red) and stationary (blue) particles at times (a) $t = 1000$, (b) $t = 2000$, (c) $t = 4000$ and  (d) $t = 8000$. Energetic particles are 
injected at the center. All collisions are inelastic with $r=0.1$. The data are for the conserved model.}
\label{fig:snapshotsinelastic}
\end{figure}

We look for scaling solutions, similar to that found for the problem with a single impact in Ref.~\cite{zjabeen2010epl}. Let $R_t$ be the typical radius of the disturbance at time $t$. We assume that it is the only relevant length scale in the problem. We assume a power-law growth for the radius of disturbance, $R_t\sim t^{\alpha}$. The typical velocity $v_t$ is then given by, $v_t \sim dR/dt \sim t^{\alpha-1}$. The total number of moving particles that have undergone collisions $N_t$ is given by the volume swept out by the disturbance in the conserved model, and the sum of the volume swept out by the disturbance and
the injected particles for the nonconserved model. The volume swept out by the disturbance scales as $R_t^{d} \sim t^{\alpha d}$, where $d$ is the spatial dimension, while the number of injected particles scales as $J t$. Therefore, in the limit of large time, $N_t\sim R_t^{d} \sim t^{\alpha d}$ for the conserved
model and $N_t\sim R_t^{d} \sim t^{\max[\alpha d,t]}$ for the nonconserved model. We discuss the two models separately.

\subsection{Conserved model \label{sec:casea}}

The energy of the system scales as
\begin{equation}
E_t\sim N_t v_t^2 \sim t^{\alpha(d+2)-2}.
\end{equation}
The exponent $\alpha$ may be determined for the elastic and inelastic cases using different conservation laws. For the elastic system, 
energy is not dissipated during collisions. However, due to the constant driving, the total energy must increase linearly with time, i.e., $E_t \sim t$ . Comparing it with the scaling behavior of energy $E_t\sim  t^{\alpha(d+2)-2}$, we conclude
\begin{equation}
 \alpha = \frac{3}{d+2}, \quad r=1.
 \label{eq:elastic}
\end{equation}
This result coincides with the power-law scaling exponent  obtained in the case of astrophysical blast waves~\cite{jpostriker1988rmp}.

For the inelastic system, the total energy is no longer conserved. However, the formation of the bands, as can be seen in Fig.~\ref{fig:snapshotsinelastic}, implies that
there is no transfer of momentum from a point in the band to a point diametrically opposite to it by  particles streaming across. Thus, once the bands form, radial momentum is  conserved during collisions and  flows radially outward~\cite{zjabeen2010epl,snpathak2012pre}. Due to the continuous driving, the radial momentum must  increase linearly with time $t$~\footnote{Also see Section II of Ref.~\cite{snpathak2012pre} for detailed discussion about the radial momentum conservation when there is no driving}. We confirm this in simulations by measuring radial momentum as  the sum of the radial velocities of all the moving particles. As shown in  Fig.~\ref{fig:radmomentum}, radial momentum increases linearly with time, at large times, in  both two and three dimensions. There is an initial transient period (see the inset of Fig.~\ref{fig:radmomentum}), where the initial growth is not linear, reflecting the time taken to form  stable dense
bands. The radial momentum, in terms of the exponent $\alpha$, scales as $N_t v_t \sim t^{\alpha (d+1)-1}$. Comparing it with the linear increase in $t$, we obtain
\begin{equation}
 \alpha = \frac{2}{d+1}, \quad r ~\textless ~1.
 \label{eq:inelastic}
\end{equation}
\label{sec:scaling}
\begin{figure}
\includegraphics[width=\columnwidth]{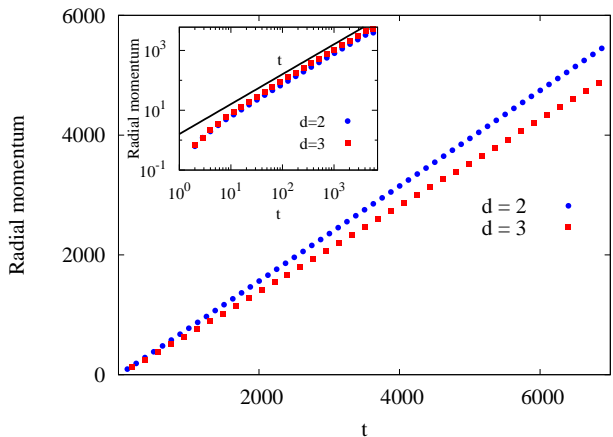}
\caption{(Color online) Radial momentum as a function of time $t$ for two and three-dimensional inelastic systems, showing a linear increase. The inset shows the data on a log-log scale, which show an initial transient regime before the linear growth is attained. The data are for the conserved model.
}
\label{fig:radmomentum}
\end{figure}

\subsection{Non-conserved model}

We show that the non-conserved model has the same scaling laws as described in Eqs.~(\ref{eq:elastic}) and (\ref{eq:inelastic}).
The energy of the system scales as
\begin{equation}
E_t\sim N_t v_t^2 \sim t^{\max[\alpha d,1]+2\alpha -2}.
\label{eq:6}
\end{equation}
In the elastic case, energy is conserved and $E_t \sim t$. Comparing with Eq.~(\ref{eq:6}), we obtain $\alpha=3/(d+2)$ if $\alpha d \geq1$ and
$\alpha=1$ if $\alpha d <1$. For $d \geq 1$, the only solution is $\alpha=3/(d+2)$, as obtained for the conserved model [see Eq.~(\ref{eq:elastic})].

For the inelastic case, the radial momentum increases linearly with time (see Sec.~\ref{sec:casea}). 
The radial momentum scales as $N_t v_t \sim t^{\max[\alpha d,1] +\alpha-1}$. Comparing it with the linear increase in $t$, we obtain
$\alpha=2/(d+1)$ if $\alpha d \geq1$ and
$\alpha=1$ if $\alpha d <1$. For $d \geq 1$, the only solution is $\alpha=2/(d+1)$, as obtained for the conserved model [see Eq.~(\ref{eq:inelastic})].

We conclude that the scaling laws are identical for both the conserved and non-conserved models. In the remaining part of the paper, we 
discuss only the conserved model.

\section{\label{numericresult} Numerical Results}

All the numerical results presented in this section are for the conserved model. The results for the nonconserved model are similar
and omitted for the sake of brevity.

\subsection{\label{elastic}Elastic}

We first show that the power-law growth of the shock radius $R_t$, the number of moving particles $N_t$, and the total energy $E_t$, as obtained
in Sec.~\ref{scalingargument} using scaling arguments,  is correct, using event-driven molecular dynamics simulations. For the elastic system, the scaling arguments
predict $R_t \sim t^{3/4}$, $E_t\sim t$, and $N_t \sim t^{3/2}$ in two dimensions and  $R_t \sim t^{3/5}$, $E_t\sim t$, and $N_t \sim t^{9/5}$ in three dimensions. 
The results from  simulations, shown in Figs.~\ref{fig:numericalresultelastic}(a)--\ref{fig:numericalresultelastic}(c) for $R_t$, $E_t$, and   $N_t$, respectively, are in excellent
agreement with the above scaling and confirm the value of
the exponent $\alpha$ as given by Eq.~(\ref{eq:elastic}).
\begin{figure}
\includegraphics[width=\columnwidth]{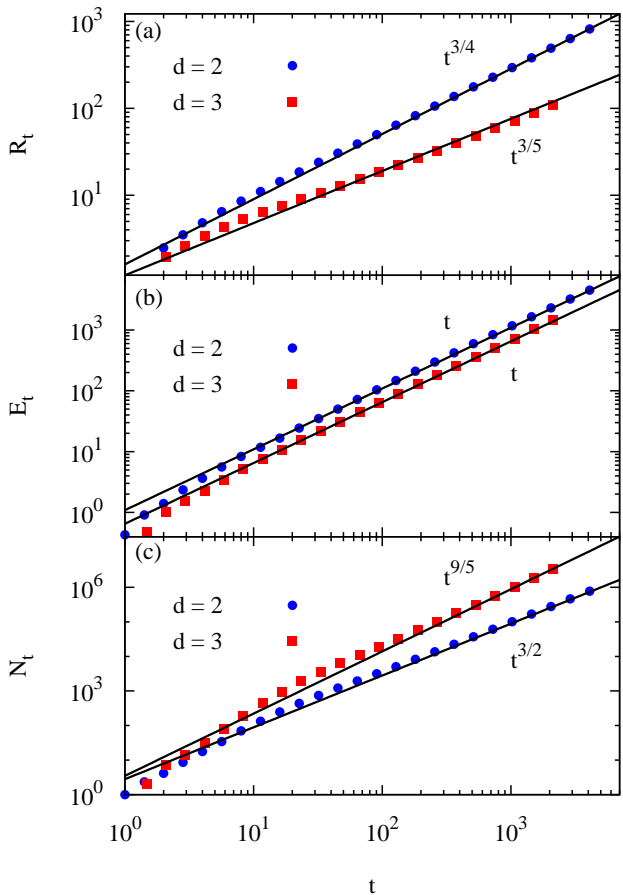}
\caption{(Color online) Simulation results for the elastic system ($r=1$) for the temporal variation of (a) radius $R_t$, (b) kinetic energy $E_t$, and (c) number of moving particles $N_t$ in two and three dimensions. The solid lines are power laws with exponents as predicted by the 
scaling arguments presented in the text. The data are for the conserved model.}
\label{fig:numericalresultelastic}
\end{figure}

The scaling argument leading to the exponent in Eq.~(\ref{eq:elastic}) assumes the existence of only one length  and one velocity scale, and leads to
the correct scaling of the bulk quantities  $R_t$, $N_t$, and $E_t$ with time. 
This assumption may be  further checked by studying the scaling  behavior of  local space-dependent physical quantities.  
We define 
coarse-grained radial density distribution function $\rho(r,t)$ as the number of moving particles per unit volume, located within a shell of radius of 
$r$ to $r+dr$. Similarly, the radial velocity distribution function $v(r,t)$ and the radial energy distribution function $e(r,t)$ are defined as the average radial velocity of particles and the average kinetic energy per unit volume, respectively, contained within the shell at any time $t$. 
We expect these local coarse-grained quantities to have the following scaling forms:
\bea 
\rho(r,t)&\sim& f_{\rho}(r/t^\alpha), \nonumber \\
v(r,t)&\sim& t^{\alpha -1} f_v(r/t^\alpha), \nonumber \\
e(r,t)&\sim& t^{-\beta}f_e(r/t^\alpha),
\label{dfs}
\eea 
where $\beta = 2 (1 - \alpha)$, since $e$ scales as $v^2$.    
\begin{figure}
\includegraphics[width=\columnwidth]{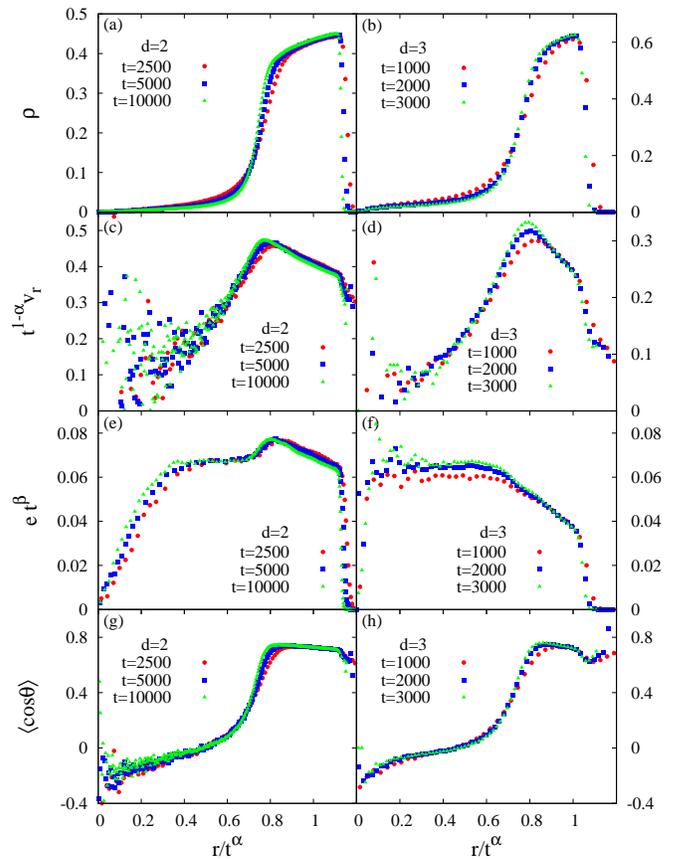}
\caption{(Color online)  Scaled radial distribution functions against scaled distances $r/t^{\alpha}$ for the elastic gas: (a) $\rho(r,t)$, (c) $v(r,t)$, (e) $e(r,t)$, and (g) $\langle \cos \theta (r,t)\rangle$  in two dimensions and (b), (d), (f), and (h) corresponding quantities in three dimensions. Here  $\alpha = 3/(d+2)$,
as in Eq.~(\ref{eq:elastic}), and $\beta= 2(1-\alpha)$. The data are for the conserved model.}
\label{fig:distributionfunctionselastic}
\end{figure}

In Figs.~\ref{fig:distributionfunctionselastic}(a) and \ref{fig:distributionfunctionselastic}(b), when  $\rho(r,t)$ for the elastic system is plotted against the scaled distance $r/t^{\alpha}$,  the data for different times collapse onto a single curve for $\alpha=3/4$ in two dimensions and for $\alpha=3/5$ in three dimensions.
The curve 
reveals that there is a substantial number of moving particles spread out between the location of the  shock front  (around scaled distance $r/t^{\alpha} \approx 1$) and scaled distances approximately equal to $0.5$.  However, the curve is nonzero and decreases to zero (as a power law) for small distances. 
Thus the region of disturbed particles does not have  an empty core, unlike the case of the inelastic system, as we will see below.  
From Figs.~\ref{fig:distributionfunctionselastic}(c)--\ref{fig:distributionfunctionselastic}(f)  we observe  that data for $v(r,t)$ and $e(r,t)$ also collapse onto a single curve in both two and
three dimensions when 
scaled as in Eq.~(\ref{dfs}) with 
the same values of $\alpha$.  Both radial velocity and density initially increase as the distance from the shock front increases. This leads to more 
compaction near the shock front due to faster particles pushing against the slower particles.  
Finally, in order to understand better the direction of motion of the particles in this driven gas,  we calculate the distribution function of  
$\langle \cos \theta (r,t)\rangle$, where $\theta$ is the angle made by the instantaneous particle velocity with respect to the 
outward unit radial  vector at its location, and the averaging is performed over all particles contained within the shell from radius $r$ to $r+dr$. 
In Figs.~\ref{fig:distributionfunctionselastic}(g) and \ref{fig:distributionfunctionselastic}(h) we see that for 
small values of the scaled distance less than $0.2$ the scaling function is negative, while for scaled distances greater than$0.8$, its value is positive and close to $1$, for both two
and three dimensions. This implies that near the shock front the particles are mostly directed radially outward, while near 
the center of the sphere the particles are on average moving inward, a feature related to the fact that the particle collisions are elastic.  The inward-moving particles are responsible for the transfer of radial momentum across the origin and lead to the breakdown of conservation of  radial momentum in a
particular direction.

\subsection{Inelastic}

Now we turn to the case more relevant to granular matter, namely, of systems with particles suffering {\it inelastic} collisions. The 
scaling dependence on time $t$ of various quantities in such systems relies on the basic assumption of radial momentum growing linearly as a function of time $t$ (see Sec.~\ref{scalingargument}). In Fig.~\ref{fig:snapshotsinelastic} we saw that the perturbed particles cluster in an outward moving narrow band.  
For the inelastic system, the scaling arguments
predict $R_t \sim t^{2/3}$, $E_t\sim t^{2/3}$, and $N_t \sim t^{4/3}$ in two dimensions and  $R_t \sim t^{1/2}$, $E_t\sim t^{1/2}$, and $N_t \sim t^{3/2}$ in three dimensions.
The results from  simulations, shown in Fig.~\ref{fig:numericalresultinelastic}(a)--\ref{fig:numericalresultinelastic}(c) for $R_t$, $E_t$, and   $N_t$, respectively, are in excellent
agreement with the above scaling and confirm the value of
the exponent $\alpha$ as given by Eq.~(\ref{eq:inelastic}).
\begin{figure}
\includegraphics[width=\columnwidth]{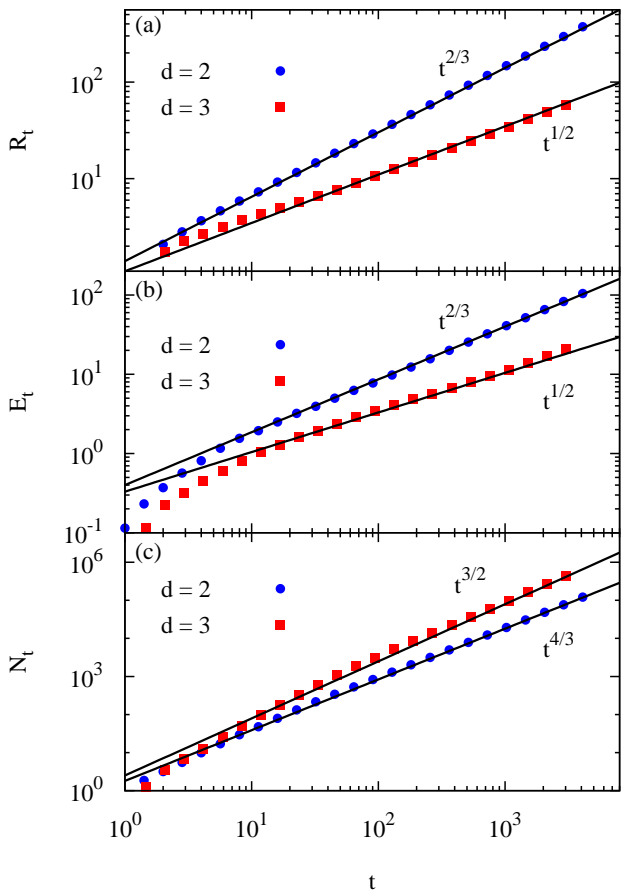}
\caption{(Color online) Simulation results for the inelastic system ($r=0.1$) for the temporal variation of (a) radius $R_t$, (b) kinetic energy $E_t$, and (c) number of moving particles $N_t$ in two and three dimensions. The solid lines are power laws with exponents as predicted by the 
scaling arguments presented in the text. The data are for the conserved model.}
\label{fig:numericalresultinelastic}
\end{figure}

Next we study the radial distribution functions for the inelastic gas and compare them with the elastic cases considered in Sec.~\ref{elastic}. 
The data for the different distributions for different times collapse onto a single curve when scaled as in Eq.~(\ref{dfs}) with
$\alpha$ as in Eq.~(\ref{eq:inelastic}) for both two dimensions [see Figs.~\ref{fig:distributionfunctionsinelastic}(a), \ref{fig:distributionfunctionsinelastic}(c), \ref{fig:distributionfunctionsinelastic}(e), and \ref{fig:distributionfunctionsinelastic}(g)] and
three dimensions [see Figs.~\ref{fig:distributionfunctionsinelastic}(b), \ref{fig:distributionfunctionsinelastic}(d), \ref{fig:distributionfunctionsinelastic}(f), and \ref{fig:distributionfunctionsinelastic}(h)].
From Figs.~\ref{fig:distributionfunctionsinelastic}(a) and \ref{fig:distributionfunctionsinelastic}(b) we see that the particle density is  highly localized between scaled distances $0.8$ and $1$ and falls to
zero rapidly for smaller scaled distances; this is to be compared to  the elastic gases [see Figs.~\ref{fig:distributionfunctionselastic}(a) and \ref{fig:distributionfunctionselastic}(b)] where
there is a larger spatial spread of density. Similar spatial localization is also observed in the velocity and energy distribution functions [see  
Fig.~\ref{fig:distributionfunctionsinelastic}(c)--\ref{fig:distributionfunctionsinelastic}(f)].  Another 
clear indication of the narrow banding of inelastic particles moving nearly perfectly radially outward is that the distribution $\langle \cos \theta (r,t)\rangle$ approaches  the value $1$ [see Figs.~\ref{fig:distributionfunctionsinelastic}(g) and \ref{fig:distributionfunctionsinelastic}(h)].  Like for the elastic case, the radial velocity increases as one
moves away from the shock front, stabilizing the dense bands containing the particles. 
\begin{figure}
\includegraphics[width=\columnwidth]{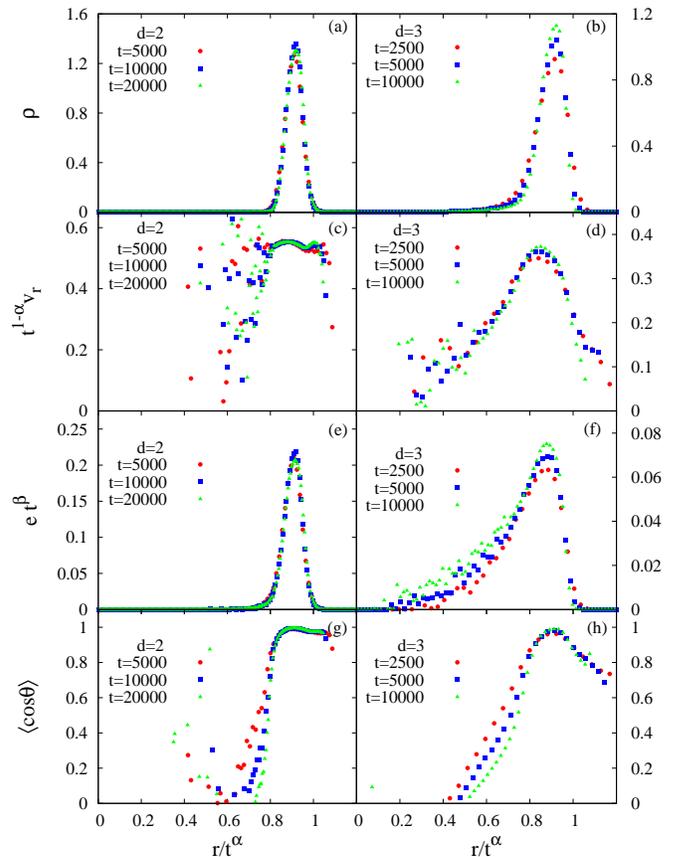}
\caption{(Color online)   Scaled radial distribution functions against scaled distances $r/t^{\alpha}$ for the inelastic gas: (a) $\rho(r,t)$, (c) $v(r,t)$, (e) $e(r,t)$, and (g) $\langle \cos \theta (r,t)\rangle$  in two dimensions and (b), (d), (f), and (h) corresponding quantities in three dimensions.  Here $\alpha = 2/(d+1)$,
as in Eq.~(\ref{eq:inelastic}), and $\beta= 2(1-\alpha)$. The data are for the conserved model.}
\label{fig:distributionfunctionsinelastic}
\end{figure}

\section{\label{comparisonwithexperiments} Comparison with experiments}

There are quite a few experiments~\cite{xcheng2008natphy,bsandnes2007prl,sfpinto2007prl,ojohnsen2006pre,hhuang2012prl} that study pattern formation in a layer of 
granular matter driven  locally at the center through the  injection of another material, gas or liquid,
but not all of them study physical quantities, which is  relevant for the predictions of this paper.  In this section we discuss two experiments that provide quantitative data on driven granular particles and we show how our scaling theory and simulations provide an  explanation for the radial growth law as seen in these experiments.

The first experiment of interest is pattern formation in  spherical glass beads that are distributed uniformly within a circular Hele-Shaw cell~\cite{xcheng2008natphy}. The beads, initially at rest, were  perturbed by the continuous injection of pressurized nitrogen  through a hole at the 
center of the bottom plate of the cell. The driving was uniform (similar to what we assume in this work). The cell boundary was open so that any
bead driven to the edge could freely flow out of the cell.   The patterns formed were recorded with high-speed camera. 
When the driving pressure was high enough, the continuous perturbation led to the formation of a time-dependent growing 
viscous fingering pattern.  Our interest is the 
radial growth law of this pattern in the early stage; at the late stage beyond some characteristic time $t_0$,  a wild growth in radius due to
effect of boundaries is seen, which is not of interest in this paper.  We replot the published data [Fig.~1(c) in Ref.~\cite{xcheng2008natphy}] in 
Fig.~\ref{fig:naturesnapshot} for scaled radius $R/R_0$ against scaled time $t/t_0$, where $R_0 = R(t_0)$.  
Quite strikingly, we find that the data converge close to the power law  $R_t \sim t^{2/3}$, as shown in Fig.~\ref{fig:naturesnapshot}, consistent with our
theoretical prediction for the two-dimensional inelastic system [see Eq.~(\ref{eq:inelastic})]. However, the scaling analysis assumes that the only means of dissipation is inelasticity.   The experiment has dissipative frictional forces too, but it is evident from the data being consistent with the power law that possibly the frictional effect is 
nullified by the critical pressure, beyond which beads start moving,  and eventually inelasticity remains as the dominant mechanism of dissipation.  We note  that the experimental  paper~\cite{xcheng2008natphy} erroneously mentions a  linear growth of radius, but it is clear that the line proportional to $t$ in  Fig.~\ref{fig:naturesnapshot}  describes the data poorly. We also note that the  power law $t^{1/2}$ in  Fig.~\ref{fig:naturesnapshot} is a poorer
fit to the data than the  power law $t^{2/3}$.
\begin{figure}
\includegraphics[width=\columnwidth]{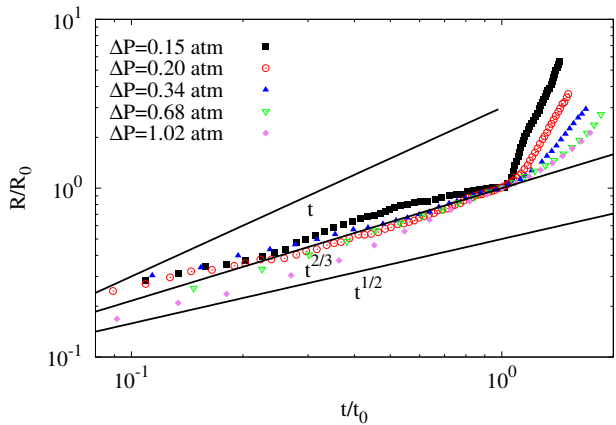}
\caption{(Color online) Experimental data (taken from Ref.~\cite{xcheng2008natphy}) 
for the scaled radius $R$ of the longest finger from the center, as a function of normalized time $t/t_0$.
Here $R_0=R(t_0)$. The data have been plotted for different gas overpressures. 
The solid lines are power laws $t^{1/2}$, $t^{2/3}$, and $ t^1$ and are  shown for reference.}
\label{fig:naturesnapshot}
\end{figure}

We look at another similar experiment with granular material confined in a circular Hele-Shaw cell with central air injection~\cite{ojohnsen2006pre}. 
When the injection pressure is sufficient enough, the particles in the system move out by forming a central (roughly circular) region devoid of particles. Around this central region, there is a zone where the granular material is compacted. The patterns formed have been recorded by using a high-speed, high-resolution CCD camera. The data obtained from this experiment [Fig.~13(a) in Ref.~\cite{ojohnsen2006pre}] also follow the power law $R_t \sim t^{2/3}$
as shown in Fig.~\ref{fig:presnap}, consistent with our growth-law exponent  [see Eq.~(\ref{eq:inelastic}) with $d=2$]. We note that the  power laws $t^{1/2}$
and $t^1$  in  Fig.~\ref{fig:presnap} are poorer
fits to the data than the  power law $t^{2/3}$. Thus, again we see that the simple 
scaling law obtained  the from dominance of inelastic  dissipation,  and band formation,  is experimentally relevant. 
\begin{figure}
\includegraphics[width=\columnwidth]{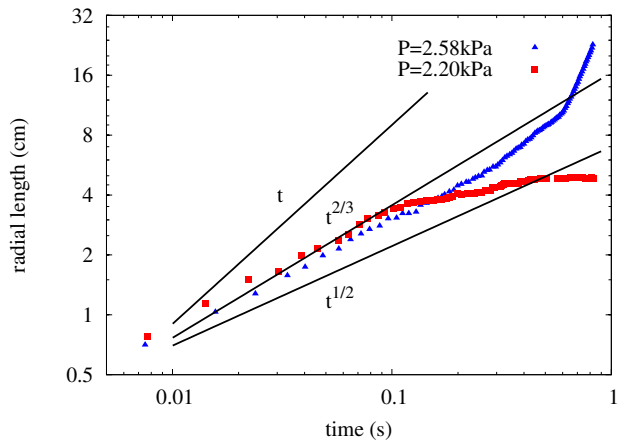}
\caption{(Color online) Experimental data (taken from Ref.~\cite{ojohnsen2006pre}) for the growth of maximum radial coordinate of the central zone of disturbance with time for two different values of injection pressures. The solid lines are power laws $t^{1/2}$, $t^{2/3}$,  and $t^1$ and are  shown for reference.}
\label{fig:presnap}
\end{figure}

\section{\label{conclusion} Conclusion and discussion}

We studied shock propagation in a granular system that is continuously driven in a  localized region. We analyzed both 
the elastic and inelastic systems  through scaling arguments and extensive event-driven molecular dynamics simulations. By identifying 
that energy grows linearly in the elastic system and radial momentum grows linearly in the inelastic system, the exponents governing
the power-law growth of bulk quantities such as radius of disturbance and number of moving particles were obtained. For the inelastic
system, the linear growth of radial momentum crucially depended on the formation of dense bands enclosing an empty region, due to inelastic collision, 
as seen in the simulations. There are very few driven granular systems where exact results can be obtained. The solution in this paper provides
an example where the exponents, presumably exact, may be determined through scaling arguments.

We analyzed two experiments on pattern formation that arise due to  the injection of a gas at localized point  in a two-dimensional granular medium. The experimentally obtained radial growth of the pattern was shown to be consistent with the results in this paper, even though the present study ignores friction that would appear to be relevant in experiments. The experimental patterns show the formation of bands that  have fractal structure, which is not captured by our model. However, the detailed structure of the bands does not play a role in determining the growth-law exponent, as the scaling arguments required only conservation of radial momentum, which in turn depends only on the existence of a band enclosing an empty region and not on its structure.

We described numerical results for the  model where the  injected energetic particles were removed from the system after their first collision. 
However, we presented scaling arguments to show that the power-law exponents for the nonconserved model, in which the  injected energetic 
particles remain in the system, are identical to that of the conserved model.
Simulations are also consistent with the predictions of scaling theory.  Such models may be valid
for experiments where granular material is driven through injection of other granular material.

Unlike the power-law exponents, it does not appear to be possible to analytically determine the form of the
scaling functions for the different local densities. For the elastic system, one might ask whether
the TvNS solution~\cite{gtaylor1950,lsedov_book,jvneumann1963cw} that describes shock propagation following an intense blast may be modified
to the case of continuous driving. The local conservation laws of density, energy, and momentum
continue to hold for  localized continuous driving away from the source. However, we find in our
preliminary studies that the solution develops singularities
at a finite distance between the origin and the shock front. This could be because the additional
assumption of local thermal equilibrium made in the TvNS solution may not hold when the driving
is continuous. A detailed analysis of the elastic case is a promising area for future study.

\begin{acknowledgments}
The simulations were carried out on the
supercomputer Nandadevi at The Institute of Mathematical Sciences.
\end{acknowledgments}


%

\end{document}